\documentclass{LMCS}
\usepackage{amssymb}
\usepackage{amsmath}
\usepackage{amsthm}
\usepackage{verbatim}
\usepackage[usenames]{color}
\usepackage{hyperref}

\newtheorem{la}[thm]{Lemma}

\theoremstyle{definition}
\newtheorem{df}[thm]{Definition}
\newtheorem{ex}[thm]{Example}   
\newtheorem{rmk}[thm]{Remark}

\newenvironment{ls}{\begin{itemize}}{\end{itemize}}
\newenvironment{lsnum}{\begin{enumerate}}{\end{enumerate}}
\newenvironment{unn}[1]{\bigskip\noindent\textbf{#1}\quad}{\par\bigskip}

\newcommand{\notarrow}{\kern .42em\not\kern -.42em\longrightarrow}

\newcommand{\DD}{\Delta^+}
\newcommand{\ans}{\dot}
\newcommand{\bbb}[1]{\ensuremath{\mathbb {#1}}}
\newcommand{\bld}[1]{\ensuremath{\mathbf {#1}}}

\newcommand{\dom}[1]{\ensuremath{{\text{Dom}}(#1)}}
\newcommand{\emp}{\varnothing} \renewcommand{\phi}{\varphi}

\newcommand{\mst}[1]{\ensuremath{\{\kern-0.33em\{#1\}\kern-0.33em\}}}
\newcommand{\ran}[1]{\ensuremath{{\text{Range}}(#1)}}
\newcommand{\restr}{\mathop{\upharpoonright}}

\newcommand{\scr}[1]{\ensuremath{\mathcal {#1}}}

\newcommand{\sq}[1]{\ensuremath{\langle#1\rangle}}

\newcommand{\ttt}[1]{\ensuremath{\mathtt {#1}}}

\renewcommand{\th}{\ensuremath{{}^{\text{th}}}}
\newcommand{\Issued}{\text{Issued}}

\newcommand{\initeq}{\unlhd}

\def\doi{3 (4:3) 2007}
\lmcsheading%
{\doi}
{1--29}
{}
{}
{Jun.~11, 2007}
{Nov.~\phantom{0}5, 2007}
{}   

\begin{document}

\title[Interactive Small-Step Algorithms I]{Interactive Small-Step
  Algorithms I: Axiomatization}

\author[A.~Blass]{Andreas Blass\rsuper a} 
\address{{\lsuper a}Mathematics Dept.,
  University of Michigan, Ann Arbor, MI 48109, U.S.A.}
\email{ablass@umich.edu}
\thanks{{\lsuper a}Blass was partially supported by NSF grant DMS--0070723 and by
  a grant from Microsoft Research.}

\author[Y.~Gurevich]{Yuri Gurevich\rsuper b}
\address{{\lsuper b}Microsoft Research, One Microsoft Way, Redmond, WA 98052,
U.S.A.}
\email{gurevich@microsoft.com}

\author[D.~Rosenzweig]{Dean Rosenzweig\rsuper c} 
\address{{\lsuper c}University of Zagreb, FSB, I. Lu\v ci\'ca 5, 10000 Zagreb,
  Croatia}
\email{dean@math.hr}
\thanks{{\lsuper c}Rosenzweig was partially supported by the grant 0120048 from
  the Croatian Ministry of Science and Technology and by Microsoft
  Research.} 

\author[B.~Rossman]{Benjamin Rossman\rsuper d}
\address{{\lsuper d}Computer Science Dept., M.I.T., Cambridge, MA 02139, U.S.A.}
\email{brossman@mit.edu}

\keywords{Interactive algorithm, small-step algorithm, abstract state
machine, abstract state machine thesis, behavioral equivalence,
ordinary algorithms, query.}
\subjclass{F.1.1, F.1.2, F.3.1}

\begin{abstract}
  In earlier work, the Abstract State Machine Thesis --- that
  arbitrary algorithms are behaviorally equivalent to abstract state
  machines --- was established for several classes of algorithms,
  including ordinary, interactive, small-step algorithms.  This was
  accomplished on the basis of axiomatizations of these classes of
  algorithms.  Here we extend the axiomatization and, in a companion
  paper, the proof, to cover interactive small-step algorithms that
  are not necessarily ordinary.  This means that the algorithms
  (1)~can complete a step without necessarily waiting for replies to
  all queries from that step and (2)~can use not only the
  environment's replies but also the order in which the replies were
  received.
\end{abstract}

\dedicatory{While this paper was being revised, we received the sad
  news of the death of our co-author,\\ Dean Rosenzweig.  We dedicate
  this paper to his memory.\\
\hfill Andreas Blass, Yuri Gurevich, Benjamin Rossman}

\maketitle


\section{Introduction}
\label{intro}

The Abstract State Machine (ASM) Thesis, first proposed in
\cite{thesis} and elaborated in \cite{tutorial,lipari}, asserts that
every algorithm is equivalent, on its natural level of abstraction, to
an abstract state machine.  Beginning in \cite{seqth} and continuing
in \cite{parth}, \cite{oa1}, \cite{oa2}, and \cite{oa3}, the thesis
has been proved for various classes of algorithms.  In each case, the
class of algorithms under consideration was defined by postulates
describing, in very general terms, the nature of the algorithms. In
each case the main theorem was that all algorithms of this class are
equivalent, in a strong sense, to appropriate ASMs. The work in
\cite{seqth,oa1,oa2,oa3} directly leading to the present paper is
briefly overviewed in the following section.

The present paper continues this tradition, but with an important
difference.  Previously, the standard syntax of ASMs, as presented in
\cite{lipari}, was adequate, with only very minor modifications.  Our
present work, however, requires a significant extension of that
syntax.  The extension allows an ASM program to refer to the order in
which the values of external functions are received from the
environment, and it allows the program to declare a step complete even
if not all external function values have been determined.

The main purpose of this paper is to extend the analysis of
interactive, small-step algorithms, begun in \cite{oa1, oa2, oa3}, by
removing the restriction to ``ordinary'' algorithms.  Here we provide
postulates and definitions describing a general notion of interactive,
small-step algorithm.  In a companion paper \cite{ga2} we also extend
the syntax and semantics of abstract state machines (ASMs) so that
non-ordinary algorithms become expressible. There we provide
\begin{ls}
  \item syntax and semantics for ASMs incorporating interaction that
  need not be ordinary,
  \item verification that ASMs satisfy the postulates, and
  \item proof that every algorithm satisfying the postulates is
  equivalent, in a strong sense, to an ASM.
\end{ls}

The algorithms considered in this paper proceed in discrete steps and
do only a bounded amount of work in each step (``small-step'') but can
interact with their environments during a step, by issuing queries and
receiving replies.

Such algorithms were analyzed in \cite{oa1, oa2, oa3}, subject to two
additional restrictions, which we expressed by the word ``ordinary.''
First, they never complete a step until they have received replies to
all the queries issued during that step. Second, the action of the
algorithm---the queries issued and the next state produced---at any
step is completely determined by the algorithm, the current state, and
the function that maps the current step's queries to the environment's
answers. In other words, this answer function is the only information
from the environment that the algorithm uses.  In particular, the
order in which the environment's answers are received has no bearing
on the computation.

In the present paper, we lift these restrictions.  We allow an algorithm
to complete a step even while some of its queries remain unanswered.  We
also allow the algorithm's actions to depend on the order in which answers
were received from the environment.  See the discussion in
Section~\ref{post} and particularly Remark~\ref{time-stamp} for a
discussion of why the order of arrival of answers is of special importance
and why it cannot be subsumed in answer functions.

It was shown in \cite{oa3} that ordinary algorithms are equivalent to ASMs
of the traditional sort, essentially as described in \cite{lipari}.  In
order to similarly characterize the more general algorithms of the present
paper, we shall, in \cite{ga2}, extend the syntax and semantics of ASMs.
In particular, we provide a way to refer to the timing of the evaluations
of external functions and a way to terminate a step while some queries
remain unanswered.  See Subsection~\ref{hist} for a discussion of the
intuitive picture of interactive algorithms that leads to these particular
extensions and indicates why they suffice.

At the referees' request, we comment briefly on the impact to be
expected from this work.  Numerous issues are involved here, which
would be better addressed in a broad discussion of the ASM thesis
rather than in this introduction, but it seems appropriate to say a
few words about future prospects.  Probably the most immediate impact
that we expect from the ASM approach is on the foundations of software
engineering.  Clarification of commonly used notions --- like state,
level of abstraction, interaction --- is long overdue.  Closer to
theoretical computer science, the ASM computation model is the most
general computation model that we know.  It supports for example
computing with structures rather than strings; see \cite{unord} for
particular examples leading to interesting theoretical results and
questions.  Finally it may be disappointing that, unlike the
Church-Turing Thesis, the ASM thesis has not been used for negative
results; will it be?  It may.  The attention of theorists is moving
gradually toward the analysis of the overall behavior of algorithms
and not only their input-output behavior.

\section{Related Work}

The previous work most closely related to the present paper is the
series of papers \cite{seqth, parth, oa1, oa2, oa3}, in which the ASM
thesis is proved for other classes of algorithms --- small-step
algorithms in \cite{seqth}, parallel algorithms in \cite{parth} (both
without external intra-step interaction, but with communication between
subprocesses in the parallel case), and ordinary interactive
small-step algorithms in \cite{oa1, oa2, oa3}.
  Also, Andreas Glausch and Wolfgang Reisig \cite{glausch, gl-re}
  adapted the postulates of \cite{seqth} to describe a restricted but
  important class of ``small-step'' distributed algorithms.
In this section, we briefly overview the work in \cite{seqth, oa1,
oa2, oa3} directly leading to the present paper, and we also briefly
discuss other work that is related to our goals and analysis, though
it applies rather different methods.

\subsection{Overview of the Behavioral Theory of Algorithms}
\label{sec:overview}

Many algorithms are naturally understood on a high level of abstraction;
for example states could be relational structures (databases).  The step
to a simulation operating over string representations of the abstract
entities may lower the abstraction level of the algorithm; see section
\ref{sec:other} for discussion.  Such a step may be less trivial than
meets the eye: for instance it is not known whether there is a
polynomial-time algorithm that, given two adjacency matrices, determines
whether they represent the same graph.

The Abstract State Machine (ASM) model of \cite{lipari} intended to
capture arbitrary algorithms at their natural levels of abstraction.  The
ASM Thesis of \cite{lipari} asserts that every algorithm is equivalent, on
its natural level of abstraction, to an abstract state machine.
Subsequent experimentation provided confirmation of the thesis
\cite{ASM,AsmL,BS}.

Paper \cite{seqth} was the first of a series of papers offering
\emph{speculative justification} for the thesis, for particular
classes of algorithms. They all follow the same general pattern;
\begin{enumerate}
\item Describe axiomatically a class \bld{A} of algorithms.
\item Define behavioral equivalence of \bld{A} algorithms.
\item Define a class \bld{M} of abstract state machines.
\item Prove the following characterization theorem for \bld{A}:
$\bld{M} \subseteq \bld{A}$, and every $A \in \bld{A}$ is
behaviorally equivalent to some $M \in \bld{M}$.
\end{enumerate}
The characterization provides a theoretical programming language for
\bld{A} and opens the way for more practical languages for
\bld{A}. The justification of the ASM Thesis thus obtained is speculative
in two ways:
\begin{itemize}
\item The claim that \bld{A} captures the intuitive class of intended
algorithms is open to criticism.
\item Definition of behavioral equivalence is open to criticism.
\end{itemize}
But the characterization of \bld{A} by \bld{M} is precise, and in this
sense the procedure \emph{proves} the ASM thesis for the class of
algorithms \bld{A} modulo the chosen behavioral equivalence.

So \bld{A} should be as broad as possible within the intended class of
algorithms, and behavioral equivalence should be as fine as possible, even
if coarser notions of equivalence might suffice for many purposes.  The
finer the behavioral equivalence, the stronger the characterization
theorem.

In this subsection we briefly overview the realization of this program
for isolated small-step algorithms in \cite{seqth} and ordinary
interactive small-step algorithms in \cite{oa1,oa2,oa3}.

\subsubsection{Isolated Small-Step
Algorithms}\label{sec:overview:seqth}

The algorithms of \cite{seqth} are executed by a single sequential agent
and are isolated in the following sense: there is no interaction with the
environment during the execution of a step.  The environment can intervene
in between algorithm's steps.  But we concentrate on step-for-step
simulation, and so inter-step interaction with the environment can be
ignored.  This class of algorithms is axiomatized by three simple
postulates.

The \textbf{Sequential Time Postulate} says that an algorithm defines a
deterministic transition system, a (not necessarily finite-state)
automaton.  More explicitly, the algorithm determines
\begin{itemize}
\item a nonempty collection of states,
\item a nonempty subcollection of initial states, and
\item a state-transition function.
\end{itemize}
The algorithm is presumed to be deterministic.  Nondeterministic choices
involve interaction with the environment; see \cite{lipari,seqth,oa1} for
discussion.  The term state is used  in a comprehensive way.  For
example, in case of a Turing machine, a state would include not only
the control state but also the head
position and the tape contents.

The postulate does not mention a notion of final state, see \cite{seqth}
for discussion.

Sequential Time already suffices to define \emph{behavioral
equivalence} of algorithms: two algorithms are behaviorally
equivalent if they have the same
\begin{itemize}
\item states,
\item initial states, and
\item state-transition functions.
\end{itemize}
This is as fine as possible, though it may be too fine for many purposes.
But, as already noted above, the finer the equivalence, the stronger the
characterization theorem.

The \textbf{Abstract State Postulate} says that
\begin{itemize}
\item
all states are first-order structures of a fixed vocabulary,
\item the transition function does not change the base set of a
state, and
\item isomorphism of structures preserves everything, which here
means states, initial states and the transition function.
\end{itemize}
It reflects the vast experience of mathematics and mathematical logic
according to which every static mathematical situation can be adequately
represented as a first-order structure.  The idea behind the second
requirement is that, even when the base set seems to increase with the
creation of new objects, those objects can be regarded as having been
already present in a ``reserve'' part of the state.  What looks like
creation is then regarded as taking an element from out of the reserve and
into the active part of the state.  (The nondeterministic choice of the
element is made by the environment.)  See \cite{lipari,seqth,oa1,oa2} and
the next section for discussion.  The idea behind the third requirement is
that all relevant state information is reflected in the vocabulary: if
your algorithm can distinguish red integers from green integers, then it
is not just about integers.

The \textbf{Bounded Exploration Postulate} expresses the idea that a
sequential algorithm (in the traditional meaning of the term)
``computes in steps of bounded complexity'' \cite{K53}.  More
explicitly, it asserts that the values of a finite set $W$ of terms
(also called expressions), that depends only on the algorithm and not
on the input or state, determine the state change (more exactly the
set of location updates) for every step; see \cite{lipari,seqth} or
the next section for precise definitions of locations and updates.

Thus, there is a uniform upper bound on the total size of locations
that an algorithm needs to explore at any state.  Bounded Exploration
is the reason why all algorithms considered, here and in the rest of
this paper, are called `small-step'.

The characterization theorem of \cite{seqth} establishes the ASM
thesis for the class \bld{A} of algorithms defined by Sequential Time,
Abstract State, and Bounded Exploration and the class \bld{M} of
machines defined by the basic ASM language of update rules, parallel
rules and conditional rules \cite{tutorial,lipari,seqth}.

While the intent in \cite{seqth} is to capture algorithms executing
steps in isolation from the environment, a degree of \emph{intra-step
  interaction} is accommodated in the ASM literature since
\cite{lipari}: (i)~using the import command to create new elements,
and (ii)~marking certain functions as \emph{external} and allowing the
environment to provide the values of external functions.  One pretends
that the interaction is inter-step.  This requires the environment to
anticipate some actions of the algorithm.  Also, in \cite{lipari},
nesting of external functions was prohibited; the first study of ASMs
with nested external functions was \cite{oa2}.  The notion of
\emph{run} determines whether \emph{inter-step} interaction is
allowed, see \cite{lipari,seqth} for precise definitions and
discussion.

\subsubsection{Ordinary Interactive Small-Step
Algorithms}\label{sec:overview:oa}

In \cite{oa1} it is argued at length why the inter-step form of
interaction cannot suffice for all modeling needs.  As a small example
take the computation of $g(f(7))$ where $f(7)$ is an external call (a
query) with argument 7, whose result $a$ is used as the argument for a
new query $g(a)$.  An attempt to model this as inter-step interaction
would force splitting the computation into substeps. But at some level
of abstraction we may want to evaluate $g(f(7))$ within a single step.
Limiting interaction to the inter-step mode would necessarily lower
the abstraction level.

Thus \cite{oa1} sets modeling \emph{intra-step} interaction as its
goal. Different forms of interaction, such as message-passing, database
queries, remote procedure calls, inputs, outputs, signals, \ldots\ all
reduce to a single universal form: a single-reply zero-or-more-arguments
not-necessarily-blocking \emph{query}.  All arguments and the reply (if
any) should be elements of the state if they are to make sense to the
algorithm.  For a formal definition of queries see \cite{oa1}; a reminder
is given in the next section.  For a detailed discussion and arguments for
the universality of the query-reply approach see \cite{oa1}.

Articles \cite{oa1,oa2,oa3} limit themselves to interactive algorithms
which are \emph{ordinary} in the sense that they obey the following two
restrictions:
\begin{enumerate}
\item the actions of the algorithm depend only on the state and the
  replies to queries, and not on other aspects, such as relative
  timing of replies, and
\item the algorithm cannot complete its step unless it has received
replies to all queries issued.
\end{enumerate}
The first restriction means that an algorithm can be seen as operating
on pairs of form $X,\alpha$ where $X$ is a state and $\alpha$ an
\emph{answer function} over $X$: a partial function mapping queries
over $X$ to their replies. The second restriction means that all
queries issued are \emph{blocking}; the algorithm cannot complete its
step without a reply. (Some uses of non-blocking, asynchronous queries
can still be modeled, by assuming that some forms of queries always
obtain a default answer. But this is an assumption on environment
behavior.)  The present paper lifts both restrictions, and thus
extends the theory to \emph{general} interactive algorithms.

Several of the postulates of \cite{oa1} cover the same ground as the
postulates of \cite{seqth}, but of course taking answer functions into
account.  The most important new postulate is the \textbf{Interaction
  Postulate}, saying that the algorithm, for each state $X$,
determines a \emph{causality relation} $\vdash_X$ between finite
answer functions and queries.  The intuition behind $\alpha \vdash_X
q$ is this: if, over state $X$, the environment behaves according to
answer function $\alpha$ then the algorithm issues $q$.  The causality
relation is an abstract representation of potential interaction of the
algorithm with the environment.

But not all answer functions can actually arise in such interaction.  A
\emph{context} for $X,\vdash_X$ is a minimal (with respect to graph
inclusion) answer function $\alpha$ with the following closure property:
if $\beta \vdash_X q$ for some subfunction $\beta \subseteq \alpha$, then
$q \in \dom{\alpha}$.  Intuitively, this means that $\alpha$ gives answers
for exactly those queries that the algorithm would ask, given $\alpha$.
This incorporates both the idea of blocking --- all queries that are
issued must be answered --- and the more basic idea that an algorithm
cannot see replies to queries that it didn't issue.  The \textbf{Updates
Postulate} asserts that, for every context $\alpha$ for $X,\vdash_X$, the
algorithm either \emph{fails} or produces the next state.

This refines the transition relation of Sequential Time of \cite{seqth}.
The possibility of explicit failure is new here; the algorithm may obtain
replies that are absurd or inconsistent from its point of
view, and it can fail in such a case.  The next state, if there is one, is
defined by an update set, which can also contain trivial updates:
``updating'' a location to the old value.  Trivial updates do not
contribute to the next state, but in composition with other algorithms can
contribute to a clash, see \cite{oa1} and also the next section for
discussion.

The inductive character of the context definition is unwound and analyzed
in detail in \cite{oa1}. The answer functions which can occur as stages in
the inductive construction of contexts are called
\emph{well-founded}. This captures the intuition of answer functions which
can actually arise as records of interaction of an algorithm and its
environment. Two causality relations (over the same state) are
\emph{equivalent} if they make the same answer functions well-founded.
Equivalent causality relations have the same contexts but the converse is
not in general true: intermediate intra-step behavior matters.

Two ordinary interactive algorithms are \emph{behaviorally
equivalent} if
\begin{itemize}
\item they have the same states and initial states,
\item at all states they have equivalent causality relations, and
\item at all state-context pairs, either they both fail or they produce
the same update sets as required by Updates.
\end{itemize}

The \textbf{Bounded Work Postulate} of \cite{oa1} extends Bounded
Exploration of \cite{seqth} to queries.  As a consequence every
well-founded answer function is finite.  Furthermore, there is a
uniform bound on the size of well-founded answer functions.

Queries can be expressed by external functions \cite{lipari}, but there is
a complication.  In some cases different textual occurrences of the same
external function symbol applied to the same arguments can be reasonably
assumed to define the same query, and thus to represent the same reply. In
some cases, for instance with object-creation operators of OOP languages,
different occurrences of the same external function symbol with the same
arguments are always assumed to represent different queries which always
obtain different fresh replies. There are also many intermediate
possibilities. Most of \cite{oa2} is devoted to a discussion of and a
general solution to this problem using the device of
\emph{query-templates}, which we reuse here. In \cite{oa2} the syntax and
semantics of ordinary interactive ASMs is defined. The syntax is
essentially the same as the ASM language of \cite{seqth} with the addition
of external function symbols, a subset of the ASM language of
\cite{lipari}. What is new is that the restriction of \cite{lipari},
forbidding nesting of external function symbols, can now be safely lifted:
the query semantics and the analysis of causality provided by \cite{oa1}
did not exist at the time \cite{lipari} was written. Thus the semantics of
\cite{lipari} was constrained to inter-step interaction only.

The characterization theorem for the class \bld{A} of algorithms
defined in \cite{oa1} and the class \bld{M} of machines defined in
\cite{oa2} is proved in \cite{oa3}.

\subsection{Other Related Work}
\label{sec:other}

Efforts to characterize algorithms mathematically began with the work
of Alonzo Church \cite{church} and Alan Turing \cite{turing} in the
1930's.  Turing gave a careful analysis and a gradual simplification
of what can happen during an algorithmic computation performed by a
person.  The result was the Turing machine model of computation and a
mathematical characterization, now universally accepted, of the
functions (from finite strings to finite strings) that are computable
in principle, disregarding limitations of time and memory.  The
simplifications, however, typically lower the level of abstraction in
a drastic way; just compare any algorithm in \cite{alg-text1} or
\cite{alg-text2} with a Turing machine implementation of that
algorithm.  The latter involves details --- the data representation on
the tape, indications how the read/write head should find the right
place for the next ``real'' work, etc. --- that are not essential to
the algorithm (but essential for the Turing machine).  Thus, although
Turing's work successfully characterized the (in principle) computable
functions, it characterized the algorithms that compute them only to
the extent that one ignores levels of abstraction.  (``In principle''
means that there are no bounds on resources, so ``computable'' does
not mean that we are able carry out the computation in the real
world.)

The ASM thesis, in contrast, asserts that algorithms can be described at
their natural levels of abstraction by ASMs.  The textbook algorithms can
be written as ASMs without introducing irrelevant details of
implementation.  A detailed implementation, such as a Turing machine or an
assembly language program, can also be written as an ASM.

There are some other differences between the Turing machine model and the
ASM model related to lowering of the abstraction level in the Turing
machine model.  In the Turing model, inputs and outputs are strings.  ASMs
deal naturally with computations involving abstract entities.  Examples
include the Gauss elimination algorithm over arbitrary fields and the
ruler-and-compass constructions of geometry.  Particularly important
examples include computations with finite structures, e.g.\ relational
databases.  Databases can be implemented as strings.  To do that, arbitrary
choices may be necessary (to order the underlying sets).  Thus string
encoding exposes the implementation while database queries are supposed to
be implementation independent.  Furthermore, even when an algorithm works
with strings, its parallel or interactive character can be lost in a
Turing machine simulation.

Slightly before Turing's paper \cite{turing}, Church \cite{church}
proposed another mathematical characterization of computability, in
terms of definability in lambda-calculus.  Church's proposal
turned out to be equivalent to Turing's; that is, the lambda-definable
functions are just those that are computable by Turing machines.
Unlike Turing's proposal, however, Church's was not accompanied by an
attempt to analyze the general notion of algorithm; it therefore lies
considerably farther from our work.

Andrei N. Kolmogorov \cite{K53} has given another mathematical description
of computation, presumably motivated by the physics of computation rather
than by an analysis of the actions of a human computer.  For a detailed
presentation of Kolmogorov's approach, see \cite{KU58}.  Also see
\cite{abs} and the references there for information about research on
pointer machines.  Like Turing's model, these computation models
also lower the abstraction level of algorithms.

Yiannis Moschovakis \cite{mosch} proposed that the informal notion of
algorithm be identified with the formal notion of \emph{recursor}.  A
recursor is a monotone operator over partial functions whose least fixed
point includes (as one component) the function that the algorithm
computes.  The approach does not seem to scale to algorithms interacting
with an unknown environment.  See \cite[Section~4.3]{abs} for a critique
of Moschovakis's computation model.

An approach to interactive computing was pioneered by Peter Wegner and
developed in particular in \cite{GSAS}.  The approach is based on
special interactive variants of Turing machines called \emph{persistent
Turing machines}, in short PTMs.  Interactive ASMs can step-for-step
simulate PTMs.  Goldin and Wegner assert that ``any sequential interactive
computation can be performed by a persistent Turing machine'' \cite{gsw}.
But this is not so if one intends to preserve the abstraction level of the
given interactive algorithm.  In particular, PTMs cannot step-for step
simulate interactive ASMs \cite{ia05}.

The topic of the present paper is intra-step interaction between a
single sequential-time algorithm and its environment (that is,
the rest of the world).  We are not aware of any previous literature
on intra-step interaction.  On the other hand, the communication
mechanism that emerges from our study is not limited to intra-step
interaction.  Our investigation did not happen in a vacuum.  It is
hard for us to indicate particular sources that influenced our
thinking, but surely --- consciously or unconsciously --- our thinking
was influenced by the theory of distributed algorithms, by programming
languages attempting to provide means for communication between
computing agents, and by computing technology.  A few references that
seem relevant to us are \cite{bcc,lynch, platt, rr, sw}.

\section{Postulates for Algorithms}
\label{post}

This section is devoted to the description of interactive,
small-step algorithms by means of suitable definitions and
postulates.  Some parts of this material are essentially the same as
in \cite{seqth}, which dealt with small-step algorithms that
interact with the environment only between steps; some other parts
are as in \cite{oa1}, which dealt with small-step algorithms that
interact with the environment within steps but only in the
``ordinary'' manner described in the introduction.  We shall present
these parts again here, without repeating the explanations and
motivations from \cite{seqth} and \cite{oa1}, but see section
\ref{sec:overview}. For the genuinely new material, dealing with the
non-ordinary aspects of our algorithms' interaction with the
environment, we shall present not only the definitions and
postulates but also the reasons and intuitions that lie behind them.

In the following presentation of the postulates and their intuitive
motivation, we use the word ``algorithm'' (or ``interactive,
small-step algorithm'') in an informal sense.  Our discussion amounts
to an analysis of this intuitive notion, and our postulates are a
distillation of the results of the analysis.  Afterward, in
Definition~\ref{alg-def}, we give the phrase ``interactive small-step
algorithm'' a formal meaning, based on the postulates.  Thus,
Definition~\ref{alg-def} implicitly asserts a thesis, namely that the
postulates and the formal notion defined from them adequately capture
the intuitive notion that we analyzed.  In \cite{ga2}, we shall show
that algorithms in our formal sense are behaviorally equivalent to
ASMs.  In the light of this result, the thesis implicit in
Definition~\ref{alg-def} is really the ASM thesis for interactive
small-step algorithms.

Throughout the definitions and postulates that follow, we consider a
fixed algorithm $A$.  We may occasionally refer to it explicitly, for
example to say that something depends only on $A$, but usually we
leave it implicit.

\subsection{States and vocabularies}

Our first postulate is a cosmetic improvement of the State Postulate of
\cite{oa1}, and most of it is assembled from parts of postulates in
\cite{seqth}.  We refer the reader to \cite{seqth} for a careful
discussion of the first three parts of the postulate and to \cite{oa1} for
the last part.

\begin{unn}{States Postulate:} The algorithm determines
\begin{ls}
\item a finite vocabulary $\Upsilon$,
\item a nonempty set \scr S of
\emph{states} which are all $\Upsilon$-structures,
\item a nonempty subset $\scr I\subseteq\scr S$ of \emph{initial
  states},
\item a finite set $\Lambda$ of \emph{labels} (to be used in forming
  queries). 
\end{ls}
\end{unn}

As in the cited earlier papers, we use the following conventions
concerning vocabularies and structures.

\begin{conv}   \label{vocab}
\mbox{}
  \begin{ls}
    \item A vocabulary $\Upsilon$ consists of function symbols with
    specified arities.
    \item Some of the symbols in $\Upsilon$ may be marked as
    \emph{static}, and some may be marked as \emph{relational}.
    Symbols not marked as static are called \emph{dynamic}.
    \item Among the symbols in $\Upsilon$ are the logic names: nullary
    symbols \ttt{true}, \ttt{false}, and \ttt{undef}; unary
    \ttt{Boole}; binary equality; and the usual propositional
    connectives.  All of these are static and all but \ttt{undef} are
    relational.
  \item An $\Upsilon$-structure consists of a nonempty base set and
    interpretations of all the function symbols as functions on that
    base set.
  \item In any $\Upsilon$-structure, the interpretations of \ttt{true},
    \ttt{false}, and \ttt{undef} are distinct.
    \item In any $\Upsilon$-structure, the interpretations of relational
    symbols are functions whose values lie in 
    $\{\ttt{true}_X,\ttt{false}_X\}$.
  \item In any $\Upsilon$-structure $X$, the interpretation of
    \ttt{Boole} maps $\ttt{true}_X$ and $\ttt{false}_X$ to
    $\ttt{true}_X$ and everything else to $\ttt{false}_X$.
    \item In any $\Upsilon$-structure $X$, the interpretation of
    equality maps pairs of equal elements to $\ttt{true}_X$ and all
    other pairs to $\ttt{false}_X$.
    \item In any $\Upsilon$-structure $X$, the propositional connectives
    are interpreted in the usual way when their arguments are in
    $\{\ttt{true}_X,\ttt{false}_X\}$, and they take the value
    $\ttt{false}_X$ whenever any argument is not in
    $\{\ttt{true}_X,\ttt{false}_X\}$.
    \item We may use the same notation $X$ for a structure and its
    base set.
\item We may omit subscripts $X$, for example from \ttt{true} and
  \ttt{false} when there is no danger of confusion.\qed
  \end{ls}
\end{conv}

\begin{df}
A \emph{potential query} in state $X$ is a finite tuple of elements of
$X\sqcup\Lambda$.  A \emph{potential reply} in $X$ is an element of
$X$.
\qed\end{df}

Here $X\sqcup\Lambda$ means the disjoint union of $X$ and $\Lambda$.
So if they are not disjoint, then they are to be replaced by disjoint
isomorphic copies.  We shall usually not mention these isomorphisms;
that is, we write as though $X$ and $\Lambda$ were disjoint.

\subsection{Histories and interaction}  \label{hist}

Interaction of an algorithm with its environment has been analyzed at
length in \cite{oa1}.  The analysis established that all forms of
interaction reduce to queries and replies.  The algorithms of
\cite{oa1,oa2,oa3} are insensitive to the relative timing of replies
within a step, and the environment interaction during a step could be
expressed there as an answer function mapping queries to their
replies.  The following simple example shows that the expressive power of
answer functions is not sufficient in general.

\begin{ex}\label{ex:broker} For a simple example, consider a
  broker who has a block of shares to sell.  He asks two clients
  whether they want to buy the shares.  Both of them reply that they
  want to buy the whole block.  The broker will sell the shares to the
  client whose message reaches him first.
\end{ex}

\begin{df}       \label{ans-fn}
An \emph{answer function} for a state $X$ is a partial map from
potential queries to potential replies.  A \emph{history} for $X$ is a
pair $\xi=\sq{\ans\xi,\leq_\xi}$ consisting of an answer function
$\ans\xi$ together with a linear pre-order $\leq_\xi$ of its domain.
By the \emph{domain} of a history $\xi$, we mean the domain
\dom{\ans\xi} of its answer function component, which is also the
field of its pre-order component.
\qed\end{df}

Recall that a \emph{pre-order} of a set $D$ is a reflexive,
transitive, binary relation on $D$, and that it is said to be
\emph{linear} if, for all $x,y\in D$, $x\leq y$ or $y\leq x$.  The
equivalence relation defined by a pre-order is given by
$$
x\equiv y\iff x\leq y\leq x.
$$
The equivalence classes are partially ordered by
$$
[x]\leq[y]\iff x\leq y,
$$
and this partial order is linear if and only if the pre-order was.

The \emph{length} of a linear pre-order is defined to be the order
type of the induced linear ordering of equivalence classes.  (We shall
use this notion of length only in the case where the number of
equivalence classes is finite, in which case this number serves as the
length.)

We also write $x<y$ to mean $x\leq y$ and $y\not\leq x$.  (Because a
pre-order need not be antisymmetric, $x<y$ is in general a stronger
statement than the conjunction of $x\leq y$ and $x\neq y$.)  When, as
in the definition above, a pre-order is written as $\leq_\xi$, we
write the corresponding equivalence relation and strict order as
$\equiv_\xi$ and $<_\xi$.  The same applies to other subscripts and
superscripts.

We use histories to express the information received by the algorithm
from its environment during a step.  The notion of answer function
comes from \cite{oa1}, where these functions, telling which queries
have received which replies, represented the whole influence of the
environment on the algorithm's work.  When algorithms are not required
to be ordinary, then additional information from the environment,
namely the relative order in which replies were received, becomes
relevant to the computation.  This information is represented by the
pre-order part of a history.  If $p,q\in\dom{\ans\xi}$ and $p<_\xi q$, this
means that the answer ${\ans\xi}(p)$ to $p$ was received strictly
before the answer ${\ans\xi}(q)$ to $q$.  If $p\equiv_\xi q$, this
means that the two answers were received simultaneously.

The rest of this subsection is devoted to explaining in more detail
the intuition behind this formalization of the intuitive notion of
the history of an algorithm's interaction with its environment
during a step.  We do not, however, repeat here Sections~2 and 4 of
\cite{oa1}. The first of these two sections explains in detail our
reasons for using queries and replies as our model of the
interaction between an algorithm and its environment.  The second
explains the reasons for our specific definitions of (potential)
queries and replies.  Here, we presuppose these explanations and
adopt the query-reply paradigm of \cite{oa1}; see also section
\ref{sec:overview}. Our task now is to explain what is added to the
picture from \cite{oa1} when we remove the restriction to ordinary
algorithms.

Much of the information provided by an environment can and should be
viewed as being part of its replies to queries.  This includes not
only the information explicitly requested by the query but also such
varied information as ``how hard did the user (who is part of the
environment) bang on the keyboard when typing this input'' or ``at
what time was this input provided'' if such information is relevant to
the algorithm's execution.  Thus, we can view such information as being
included in the answer function ${\ans\xi}$, without referring to the second
component of a history, the pre-order.

The purpose of the pre-order, on the other hand, is to represent the
order in which the algorithm becomes aware of the environment's
answers.  Even if the environment provides a time-stamp as part of
each reply, this order cannot be read off from the replies.  The
algorithm may become aware of the replies in an order different from
that indicated by the time stamps. In the
broker example \ref{ex:broker}, the broker will sell the shares to the
client whose message reaches him first, even if the message from the
other client is sent earlier.

\begin{rmk} \label{time-stamp} Could the environment provide, as part
  of its replies, time stamps that tell when the algorithm becomes
  aware of the replies?  If so, could we then proceed as in
  \cite{oa1}, where algorithms get information from the environment
  only via the answer function?  In other words, could we dispense
  with the pre-order component in histories?

  The idea of a time stamp telling not when the reply was sent but
  when it is received is not as absurd as it might appear.  The
  factors affecting the time of receipt, such as the speed of
  connections over which the reply travels and the speed at which the
  algorithm is executed, are not specified by the algorithm and the
  state, so they are part of the environment, albeit a different part
  from that ordinarily involved in answering queries.  So the complete
  environment could conceivably assemble replies that include the time
  that the algorithm sees them.  Such an environment would be quite
  unusual, and of course it would have to be subject to the constraint
  that the time stamps accurately reflect at least the order in which
  replies are received.  But, if such an environment can be imagined,
  could we use it, at least for theoretical purposes, to avoid the
  need for the pre-order compnent in histories?  If so, then the
  present paper could stay closer to the work previously done in
  \cite{oa1,oa2,oa3}, and we would have been quite happy to avoid the
  extra work.

  The pre-orders are, however, needed, even for theoretical purposes.
Consider the broker example above.  Part of the broker's algorithm
tells him what to do if both clients reply.  To model that in the
style of \cite{oa1}, assuming that replies include accurate
information about their arrival times, the algorithm would have to
wait for both replies, read both time stamps, and act accordingly.
But in fact, the algorithm does not wait for both replies.  If it
receives one reply, it acts on the basis of that reply without waiting
for the other.  

In effect, this algorithm and others like it decide the relative order
of receipt of two replies as soon as one of the replies is received.
If one is received and the other isn't, then the received one is
earlier.  This common sense approach is unavailable if the timing
comparisons have to be based on reading time stamps embedded in the
replies.  Thus, even if we constrained our environments to always
provide time stamps that accurately indicate the time a reply is seen
by the algorithm, we would still need machinery like that in the
present paper to decide the relative order of two replies even if only
one has been seen.
\end{rmk}

\begin{rmk}
  Even though the pre-order $\leq_\xi$ is about the replies, the
  formal definition says that it pre-orders the domain of the answer
  function ${\ans\xi}$, i.e., the set of queries.  The reason is a
  technical one: Different queries may receive the same reply, and in
  that case the single reply could occur at several positions in the
  ordering.  Each query, on the other hand, is issued just once
  because, in accordance with the conventions of \cite{oa1}, if what
  appears to be the same query is issued repeatedly, then we regard
  the repetitions as distinct queries (since they can receive
  different replies).  Thus, the order in which replies are received
  can be conveniently represented by pre-ordering the associated
  queries.  It may be more intuitive to think of pre-ordering the set
  of pairs (query, reply), i.e., the graph of the answer
  function. This view of the situation would make no essential
  difference, since these pairs are in canonical bijection with the
  queries.
\end{rmk}

We emphasize that the timing we are concerned with here is logical
time, not physical time.  That is, it is measured by the progress of
the computation, not by an external clock.  If external, physical,
clock time is relevant, as in real-time algorithms, it would have to
be provided separately by the environment, for it is not part of the
program or of the state.  The relevant values of the physical time
could be regarded as the replies to repeated queries asking ``what
time is it?''

In particular, we regard a query as being issued by the algorithm as
soon as the information causing that query (in the sense of the
Interaction Postulate below) is available.  This viewpoint would be
incorrect in terms of physical time, for the algorithm may be slow to
issue a query even after it has the prerequisite information. But we
can regard a query as being logically available as soon as information
causing it is present.

This is why we include, in histories, only the relative ordering of
replies.  The ordering of queries relative to replies or relative to
each other is then determined.  The logical time of a query is the
same as the logical time of the last of the replies causing that
query.

Our use of pre-orders rather than partial orders, i.e., our allowing
two distinct replies to be received simultaneously, also reflects our
concern with logical rather than physical time.  One can argue that no
two events are exactly simultaneous in the physical sense (though they
could be so nearly simultaneous that one should treat them as
simultaneous), but logical simultaneity is certainly possible.  It
means merely that whenever the computation had access to one of the
replies it also had access to the other.

The linearity of the pre-ordering, i.e., the requirement that every
two replies be received either in one order or in the other or
simultaneously, formalizes the following important part of our view of
sequential-time (as opposed to distributed) algorithms.  (It is not a
postulate of our formal development but an informal principle that
underlies some of our definitions and postulates.)

\begin{unn}{One Executor Principle}
A small-step algorithm is executed by a single, sequential entity.
Even if there are (boundedly many) subprocesses running in parallel,
they all operate under the control of a single, master executor.
\end{unn}

The aspect of this principle that is relevant to the linearity of
histories can be summarized in the following informal principle, in
which the first part emphasizes the role of the master executor in the
interaction with the environment, while the second part is a
consequence of the sequentiality of the executor.

\begin{unn}{Holistic Principle}
The environment can send replies only to (the executor of) the
algorithm, not directly to parts of it; similarly it receives queries
only from the algorithm, not from parts of it.  Any two replies are
received by the algorithm either simultaneously or one after the
other.
\end{unn}

The pre-orders in our histories are intended to represent the order in
which replies are received by the master executor.  If there are
subprocesses, then the master may pass the replies (or information
derived from them) to these subprocesses at different times, but that
timing is part of the algorithm's internal operation (possibly also
influenced by the replies to other queries like ``what time is it?''),
not part of the environment interaction that histories are intended to
model.  The linearity of our pre-orders reflects the fact that they
represent the ordering as seen by a single, sequential entity; this
entity sees things in a sequence.

In more detail, we picture the execution of one step of an algorithm
as follows.  First, the algorithm (or, strictly speaking, its
executor) computes as much as it can with only the information
provided by the current state.  This part of the computation, the
first phase, will in general include issuing some queries.  Then the
algorithm pauses until it is ``awakened'' by the environment, which
has replied to some (not necessarily all) of the queries from phase~1.
The algorithm proceeds, in phase~2, to compute as much as it can using
the state, the new information from the environment, and
the information recorded during phase~1.  Then it again pauses
until the environment has provided some more replies (possibly to
queries from phase~2 and possibly to previously unanswered queries
from phase~1) and awakens the algorithm.  Then phase~3 begins, and
this pattern continues until the algorithm determines, in some
phase, that it is ready to complete the current step, either by
executing certain updates (computed during the various phases) of
its state or by failing (in which case the whole computation fails
and there is no next state).

The logical ordering of replies refers to the phases at which replies
were received.  That is, if $q_1<_\xi q_2$, this means that the reply
${\ans\xi}(q_1)$ to $q_1$ was received at an earlier phase than the reply
${\ans\xi}(q_2)$ to $q_2$.  Similarly, $q_1\equiv_\xi q_2$ means that these two
replies were received at the same phase.

\begin{rmk}
  We have assumed, in the preceding description of phases, that the
  algorithm is awakened to begin a new phase only when some new reply
  has been provided.  One may worry that this is not general enough,
  that the environment can awaken the algorithm even when no new
  replies are available.  The rest of this remark is devoted to
  discussing this issue and justifying our assumption; it can safely
  be skipped by readers not worried about such subtleties.
In such a case, it is natural to assume that the
algorithm, having determined that it has no new information with which
to advance the computation, simply resumes its pause until awakened
again.  In order for the algorithm to be a small-step algorithm, such
fruitless awakenings must happen only a bounded number of times per
step, with the bound depending only on the algorithm.  The reason is
that, even if all the algorithm does when awakened is to observe that
no new replies have arrived, this observing is work, and a small-step
algorithm can do only a bounded amount of it per step.

It seems possible, however, for an algorithm to admit a few fruitless
awakenings per step, and the results of the computation could even
depend on these awakenings.  Consider, for example, an algorithm that
works as follows.  It begins by issuing a query $q$ and pausing.  When
awakened, it outputs 1 and halts if there is a reply to $q$;
otherwise, it pauses again.  When awakened a second time, it outputs 2
if there is now a reply to $q$, and, whether or not there is a reply,
it halts.  Notice that, if $q$ receives the reply $r$, then whether
this is seen at the first or the second awakening doesn't affect the
history, which consists of the function $\{\sq{q,r}\}$ and the unique
pre-order on its domain $\{q\}$.  So the history fails to capture all
the information from the environment that is relevant to the
computation.

There are two ways to correct this discrepancy.  One, which we shall
adopt, is to regard the fruitless awakening as amounting to a reply to
an implicit query, of the form ``I'm willing to respond to
awakening.''  (For several fruitless awakenings, there would have to
be several such queries, distinguished perhaps by numerical labels.)
Now the two scenarios considered above, where $q$ has
received a reply at the first awakening or only at the second, are
distinguished in the histories, because the reply to the new, implicit
query will be simultaneous with the reply to $q$ in the one scenario
and will strictly precede the reply to $q$ in the other scenario.

An alternative approach would be to avoid introducing such implicit
queries but instead to replace the pre-order constituents of histories
by slightly more complicated objects, ``pre-orders with holes.''  The
idea is that the scenario where an answer to $q$ is available only at
the second awakening would be represented by the answer function
$\{\sq{q,r}\}$ as above, but the pre-order would be replaced with
something that says ``first there's a hole and then $q$.''
Formalizing this is actually quite easy, as long as the pre-order is
linear and the set $D$ to be pre-ordered is finite (as it will be in
the situations of interest to us).  A linear pre-order of $D$ with
holes is just a function $p$ from $D$ into the natural numbers.
For any $n$, the elements of $p^{-1}(\{n\})$ constitute the $n\th$
equivalence class, which may be empty in case of a hole.  Ordinary
pre-orders correspond to the special case where the image $p(D)$ is an
initial segment of \bbb N.

Although the approach using pre-orders with holes corresponds more
directly to intuition, we prefer the first approach, with implicit
queries, for two reasons.  First, it allows us to use the standard
terminology of pre-orders rather than introducing something new.
Second, its chief disadvantage, the need for implicit queries, is
mitigated by the fact that we need other sorts of implicit queries for
other purposes.  For example, in \cite[Section~2]{oa1}, implicit
queries modeled the algorithm's receptiveness to incoming
(unsolicited) messages and to multiple answers to the same query.
Notice that, in all cases, our implicit queries represent the
algorithm's willingness to pay attention to something provided by the
environment.
\qed\end{rmk}

We next define initial segments of histories, which will serve to
model the interaction with the environment part way through a step.

\begin{df}
Let $\leq$ be a pre-order of a set $D$.  An \emph{initial segment} of
$D$ with respect to $\leq$ is a subset $S$ of $D$ such that whenever
$x\leq y$ and $y\in S$ then $x\in S$.  An \emph{initial segment} of
$\leq$ is the restriction of $\leq$ to an initial segment of $D$ with
respect to $\leq$.  An \emph{initial segment} of a history
\sq{{\ans\xi},\leq_\xi} is a history \sq{{\ans\xi}\restr
S,\leq_\xi\restr S}, where $S$ is an initial segment of \dom{\ans\xi}\ with
respect to $\leq_\xi$.  (We use the standard notation $\restr$ for the
restriction of a function or a relation to a set.)  We write
$\eta\initeq\xi$ to mean that the history $\eta$ is an initial segment
of the history $\xi$.
\qed\end{df}

Notice that any initial segment with respect to a pre-order $\leq$ is
closed under the associated equivalence $\equiv$.  Notice also that if
$\sq{\ans\eta,\leq_\eta}\initeq\sq{\ans\xi,\leq_\xi}$ then $\leq_\eta$
is an initial segment of $\leq_\xi$.  We also point out for future
reference that, if two histories $\xi_1$ and $\xi_2$ are initial
segments of the same $\xi$, then one of $\xi_1$ and $\xi_2$ is an
initial segment of the other.

Intuitively, if \sq{{\ans\xi},\leq_\xi} is the history of an algorithm's
interaction with the environment up to a certain phase, then a proper
initial segment of this history describes the interaction up to some
earlier phase.

\begin{df}
  If $\leq$ pre-orders the set $D$ and if $q\in D$, then we define two
  associated initial segments as follows.
  \begin{align*}
    (\leq q)&=\{d\in D:d\leq q\}\\
    \hbox to 159 pt{\hfill}(<q)&=\{d\in D:d<q\}.\hbox to 158 pt{\hfill\qEd}
  \end{align*}
\end{df}

The following postulate is almost the same as the postulate of the
same name in \cite{oa1}.  The only difference is that we use histories
instead of answer functions.  This reflects the fact that the decision
to issue a query can be influenced by the timing of the replies to
previous queries.

\begin{unn}{Interaction Postulate}
For each state $X$, the algorithm determines a binary relation
  $\vdash_X$, called the \emph{causality relation}, between finite
  histories and potential queries.
\end{unn}

The intended meaning of $\xi\vdash_Xq$ is that, if the algorithm's
current state is $X$ and the history of its interaction so far (as
seen by the algorithm during the current step) is $\xi$, then it will
issue the query $q$ unless it has already done so.  When we say that
the history so far is $\xi$, we mean not only that the environment has
given the replies indicated in $\ans\xi$ in the order given by
$\leq_\xi$, but also that no other queries have been answered.  Thus,
although $\xi$ explicitly contains only positive information about the
replies received so far, it also implicitly contains the negative
information that there have been no other replies.  Of course, if
additional replies are received later, so that the new history has
$\xi$ as a proper initial segment, then $q$ is still among the issued
queries, because it was issued at the earlier time when the history
was only $\xi$.  This observation is formalized as follows.

\begin{df}
 For any state $X$ and history $\xi$, we define sets of queries
 \begin{align*}
   \Issued_X(\xi)&=
   \{q:(\exists\eta\initeq\xi)\,\eta\vdash_Xq\}\\
   \hbox to 127 pt{\hfill}
   \text{Pending}_X(\xi)&=\Issued_X(\xi)-\dom{\ans\xi}.
   \hbox to 127 pt{\hfill\qEd}
 \end{align*}
\end{df}

Thus, $\Issued_X(\xi)$ is the set of queries that have been
issued by the algorithm, in state $X$, by the time the history is
$\xi$, and $\text{Pending}_X(\xi)$ is the subset of those that have,
as yet, no replies.

\begin{rmk}   \label{descr-causes}
  We have described the causality relation in terms of detailed
  causes, histories $\xi$ that contain the algorithm's whole
  interaction with environment up to the time the caused query is
  issued.  A text describing the algorithm, either informally or in
  some programming language, would be more likely to describe causes
  in terms of partial information about the history.  For example, it
  might say ``if the query $p$ has received the reply $a$ but $p'$ has
  no reply yet, then issue $q$.''  This description would correspond
  to a large (possibly infinite) set of instances $\xi\vdash q$ of the
  causality relation, namely one instance for every history $\xi$ that
  fits the description, i.e., such that $\dot\xi(p)=a$ and
  $p'\notin\dom{\dot\xi}$.  More generally, whenever we are given a
  description of the conditions under which various queries are to be
  issued, we can similarly convert it into a causality relation; each
  of the conditions would be replaced by the set of finite histories
  that satisfy the condition.

The reverse transformation, from detailed causes representing the
whole history to finite descriptions, is not possible in general.  The
difficulty is that the finite descriptions might have to specify all
of the negative information implicit in a history, and this might be
an infinite amount of information.  In most situations, however, the
set of queries that might be issued is finite and known.  Then any
finite history $\xi$ can be converted into a finite description,
simply by adding, to the positive information explicit in $\xi$, the
negative information that there have been no replies to any other
queries among the finitely many that could be issued.

For semantical purposes, as in the present section, it seems natural
to think of causes as being the detailed histories.  From the behavior
of an algorithm, one can easily determine whether $\xi\vdash_Xq$ (at
least when $\xi$ is attainable as in Definition~\ref{att-def} below
and we ignore causes of $q$ that have proper initial segments already
causing $q$); just provide the algorithm with replies according to
$\xi$ and see whether it produces $q$ at that moment (and not
earlier).  It is not easy to determine, on the basis of the behavior,
whether $q$ is issued whenever a description, like ``$p$ has reply $a$
and $p'$ has no reply'' is satisfied, as this involves the behavior in
a large, possibly infinite number of situations.

For syntactic purposes, on the other hand, descriptions seem more
natural than detailed histories.  And indeed, the syntax of our ASMs
(in \cite{ga2}) will not involve histories directly but will involve
descriptions like ``$p$ has reply $a$ and $p'$ has no reply'' in the
guards of conditional rules. \qed\end{rmk}

Notice that there is no guarantee that
$\dom{\ans\xi}\subseteq\Issued_X(\xi)$, although this will be the case
for attainable histories (defined below).  In general, a history as
defined above may contain answers for queries that it and its initial
segments don't cause.  It may also contain answers for queries that
would be issued but only at a later phase than where the answers
appear.  In this sense, histories need not be possible records of
actual interactions, between the algorithm and its environment, under
the causality relation that is given by the algorithm.  The following
definition describes the histories that are consistent with the given
causality relation.  Informally, these are the histories where every
query in the domain has a legitimate reason, under the causality
relation, for being there.

\begin{df}
  A history $\xi$ is \emph{coherent}, with respect to a state
  $X$ or its associated causality relation $\vdash_X$, if
  \begin{lsnum}
    \item $(\forall q\in\dom{\ans\xi})\,q\in\Issued_X(\xi\restr(<q))$,
    and
    \item the linear order of the $\equiv_\xi$-equivalence classes
    induced by $\leq_\xi$ is a well-order.
  \end{lsnum}
\qed\end{df}
The first requirement in this definition, which can be equivalently
rewritten as
$$
(\forall q\in\dom{\ans\xi})(\exists\eta\initeq\xi)\,
(\eta\vdash_Xq\text{ and }q\notin\dom{\ans\eta}),
$$
says that, if a query $q$ receives an answer at some phase, then it
must have been issued on the basis of what happened in strictly earlier
phases.  In particular, the queries answered in the first phase, i.e.,
those in the first $\equiv_\xi$-class, must be caused by the empty
history.

The second requirement is needed only because we allow infinite
histories; finite linear orders are automatically well-orders.  The
purpose of the second requirement is to support the usual
understanding of ``cause'' by prohibiting an infinite regress of
causes.

\begin{rmk}
It will follow from the Bounded Work Postulate below in conjunction
with the Interaction Postulate above, that we can confine attention
to histories whose domains are finite.  Then clause 2 in the
definition of coherence would be automatically satisfied.
Nevertheless, we allow a coherent history $\xi$ to have infinite
domain and even to have infinitely many equivalence classes, with
respect to $\leq_\xi$, provided their ordering induced by $\leq_\xi$
is a well-ordering. There are two reasons for this generality.
First, it will allow us to state the Bounded Work Postulate in a
relatively weak form and then to deduce the stronger fact that the
histories we really need to consider (the attainable ones) are
finite.  Second, it avoids formally combining issues that are
conceptually separate. Finiteness is, of course, an essential aspect
of computation, especially of small-step computation, and the
Bounded Work Postulate will formalize this aspect.  But the notions
of coherent and attainable histories are conceptually independent of
finiteness, and so we do not impose finiteness in their definitions.
\qed\end{rmk}

\begin{rmk}   \label{cpt1}
  As mentioned above, the notion of coherence is intended to capture
  the idea of a history that makes sense, given the algorithm's
  causality relation.  Here we implicitly use the fact that we are
  describing an entire algorithm, not a component that works with
  other components to constitute an algorithm.  If we were dealing
  with a component $C$, working in the presence of other components,
  then it would be possible for queries to be answered without having
  been asked by $C$, simply because another component could have asked
  the query.  See \cite{composite} for a study of components.
\qed\end{rmk}

It follows immediately from the definitions that any initial segment
of a coherent history is coherent.

\begin{df}
A history $\xi$ for a state $X$ is \emph{complete} if
$\text{Pending}_X(\xi)=\emp$.
\qed\end{df}

Thus, completeness means that all queries that have been issued have
also been answered.  It can be regarded as a sort of converse to
coherence, since the latter means that all queries that have
been answered have also been issued earlier.

If a complete history has arisen in the course of a computation, then
there will be no further interaction with the environment during this
step.  No further interaction can originate with the environment,
because no queries remain to be answered.  No further interaction can
originate with the algorithm, since $\xi$ and its initial segments
don't cause any further queries.  So the algorithm must either proceed
to the next step (by updating its state) or fail.

\subsection{Completing a step}

At the end of a step the algorithm will, unless it fails, perform some
set of updates.  The next postulate will express this property of
algorithms formally.  First, we adopt the formalization of the notion of
update that was used in earlier work, starting with \cite{seqth}.

\begin{df}
A \emph{location} in a state $X$ is a pair \sq{f,\bld a} where $f$ is
a dynamic function symbol from $\Upsilon$ and $\bld a$ is a tuple of
elements of $X$, of the right length to serve as an argument for the
function $f_X$ interpreting the symbol $f$ in the state $X$.  The
\emph{value} of this location in $X$ is $f_X(\bld a)$.  An
\emph{update} for $X$ is a pair $(l,b)$ consisting of a location $l$
and an element $b$ of $X$.  An update $(l,b)$ is \emph{trivial} (in
$X$) if $b$ is the value of $l$ in $X$.  We often omit parentheses and
brackets, writing locations as \sq{f,a_1,\dots,a_n} instead of
\sq{f,\sq{a_1,\dots,a_n}} and writing updates as \sq{f,\bld a,b} or
\sq{f,a_1,\dots,a_n,b} instead of $(\sq{f,\bld a},b)$
or $(\sq{f,\sq{a_1,\dots,a_n}},b)$.
\qed\end{df}

The intended meaning of an update \sq{f,\bld a,b} is that the
interpretation of $f$ is to be changed (if necessary, i.e., if the
update is not trivial) so that its value at $\bld a$ is $b$.

Because our algorithms are not required to be ordinary, they may
finish a step without reaching a complete history.  The following
postulate says that they must finish the step when they reach a
complete history, if they have not already done so earlier.

\begin{unn}{Step Postulate --- Part A}
The algorithm determines, for each state $X$, a set $\scr F_X$ of
\emph{final histories}.  Every complete, coherent history has an
initial segment (possibly the whole history) in $\scr F_X$.
\end{unn}

Intuitively, a history is final for $X$ if, whenever it arises in the course
of a computation in $X$, the algorithm completes its step, either by
failing or by executing its updates and proceeding to the next step.

Since incoherent histories cannot arise under the algorithm's
causality relation, it will not matter which such histories are final.
We could have postulated that only coherent histories can be final,
but there is no need to do so.  It does no harm to allow incoherent
final histories, though they are irrelevant to the algorithm's
computation.  By allowing them, we may slightly simplify the
description, as algorithms subject to our postulates, of programs
written in some programming language; if the language allows one to
say ``finish the step'' then this can be directly translated into
final histories, without having to check for coherence.  Similar
comments apply to the remaining parts of the Step Postulate below,
where we require update sets or failure for coherent histories but
allow them also for incoherent ones.

\begin{df}   \label{att-def}
  A history for a state $X$ is \emph{attainable} (in $X$) if it
  is coherent and no proper initial segment of it is final.
\qed\end{df}

Since initial segments of coherent histories are coherent, it follows
that initial segments of attainable histories are attainable.

The attainable histories are those that can occur under the given
causality relation and the given choice of final histories.  That is,
not only are the queries answered in an order consistent with
$\vdash_X$ (coherence), but the history does not continue beyond where
$\scr F_X$ says it should stop.

Our earlier statements to the effect that incoherent histories don't
matter can be strengthened to say that unattainable histories don't
matter.  Thus, for example, although we allow one final history to be
a proper initial segment of another (even when both are coherent), the
longer of the two will be unattainable, so its being final is
irrelevant.

\begin{unn}{Step Postulate --- Part B}
  For each state $X$, the algorithm determines that certain histories
  \emph{succeed} and others \emph{fail}.  Every final, attainable
  history either succeeds or fails but not both.  
\end{unn}

\begin{df}
  We write $\scr F_X^+$ for the set of successful final histories and
  $\scr F_X^-$ for the set of failing final histories.
\end{df}

Non-final or unattainable histories may succeed or fail or both, but
this will be irrelevant to the behavior of the algorithm.  If we
restrict attention to attainable histories, then $\scr F_X^+$ and $\scr
F_X^-$ partition $\scr F_X$.

The intended meaning of ``succeed'' and ``fail'' is that a successful
final history is one in which the algorithm finishes its step and
performs a set of updates of its state, while a failing final history
is one in which the algorithm cannot continue --- the step ends, but
there is no next state, not even a repetition of the current state.
Such a situation can arise if the algorithm computes inconsistent
updates.  It can also arise if the environment gives inappropriate
answers to some queries.

For more about failures, see the discussion following the Update
Postulate in \cite[Section~5]{oa1}.  There is, however, one difference
between our present situation and that in \cite{oa1}.  There, the
algorithm depended upon getting answers to all its queries, so it is
reasonable to expect that an inappropriate answer always results in
failure.  In our present situation, however, the algorithm may be
willing to finish its step without getting the answer to a certain
query.  In this case, an inappropriate answer to that query need not
force failure.

When a history in $\scr F^+_X$ arises during the computation, the
algorithm should perform its updates and proceed to the next step.
We formalize this in the third and last part of the Step Postulate,
keeping in mind that only attainable histories can arise.  The
notation $\DD$ for the update set is taken from \cite{oa1}.  The
superscript $+$ refers to the fact that trivial updates can be
included in the update set.  Although they do not affect the next
state, they can affect whether clashes occur when our algorithm is
run in parallel with another\footnote{In fact in this paper and
    its sequel \cite{ga2} we
do not define parallel composition of algorithms. But it is implicit
in \cite{oa3} and will be explicit in \cite{composite}, and we wish
to maintain uniformity of the framework across algorithm classes.}.

\begin{unn}{Step Postulate --- Part C}
For each attainable history $\xi\in\scr F^+_X$ for a state $X$, the
algorithm determines an \emph{update set} $\DD(X,\xi)$, whose elements
are updates for $X$.  It also produces a \emph{next state}
$\tau(X,\xi)$, which
\begin{ls}
  \item has the same base set as $X$,
  \item has $f_{\tau(X,\xi)}(\bld a)=b$ if $\sq{f,\bld
  a,b}\in\DD(X,\xi)$, and
  \item otherwise interprets function symbols as in $X$.
\end{ls}
\end{unn}

The Step Postulate is the analog, for our present situation, of the
Update Postulate of \cite{oa1}.  There are two differences between
these postulates, reflecting the two requirements of ordinariness in
\cite{oa1} that have been lifted in the present work.

First, since we now consider algorithms that can finish a step
without waiting for all queries to be answered, we include in the
notion of algorithm the information about when to finish a step.
This is the notion of final history introduced in Part~A of the Step
Postulate. If an algorithm never finishes a step while queries are
still pending, then its final histories could be defined as being
the complete histories (or the complete, coherent histories), and
there would be no need to specify \scr F separately in the
algorithm.  The complete, coherent histories correspond to the
contexts, as defined in \cite{oa1}.

Second, because we allow an algorithm to take into account the order
in which it receives replies to its queries, decisions about failures
and updates depend not only on the answer function $\ans\xi$ but on
the whole history $\xi$, including the ordering information contained
in $\leq_\xi$.

The intuition and motivation behind the other aspects of the Step
Postulate are discussed in detail in \cite{seqth} and \cite{oa1}.

\begin{conv}
  In notations like $\scr F_X$, $\scr F^+_X$, $\scr F^-_X$, $\DD(X,\xi)$, and
  $\tau(X,\xi)$, we may omit $X$ if only one $X$ is under discussion.  We
  may also add the algorithm $A$ as a superscript if several
  algorithms are under discussion.
\qed\end{conv}

Notice that the Step Postulate requires $\DD(X,\xi)$ and $\tau(X,\xi)$
to be defined when $\xi$ is an attainable, final, successful history,
but it allows them to be defined also for other histories.  Any values
they may have for other histories, however, will not affect the
algorithm's computation.

Notice also that the next state is completely determined by the
current state and the update set.  So when describing an algorithm, we
need not explicitly describe $\tau$ if we have described $\DD$.

If $\DD(X,\xi)$ clashes, i.e., if it contains two distinct updates of
the same location, then the description of $\tau(X,\xi)$ is
contradictory, so the next state cannot exist.  Thus, if such a $\xi$
is attainable and final, it must be failing.  That is, clashes imply
failure.

\begin{ex}\label{ex;broker:re} We revisit the broker example
  \ref{ex:broker} in terms of our definitions.  Part of the broker's
  algorithm is easy to describe: the empty history causes two queries,
  $q_0(s,p,a)$ and $q_1(s,p,a)$.  Query $q_i(s,p,a)$ expresses the
  take-it-or-leave-it offer to client $i$ to buy a block
  of $a$ shares of stock $s$ at price $p$.  A history that contains a
  positive reply from just one client is final and successful and produces
  updates recording a sale to that client.  Note that, if both clients
  reply positively but at different times, then the history as an
  initial segment containing only the reply from the client who
  answered first, so the stock is sold to this client.  It is also
  easy to handle histories in which both clients answer negatively;
  such a history is successful and final but results in no updates (or
  perhaps in an update recording that these clients had the chance to
  buy the stock and turned it down).  Note that, although this
  situation represents a failure of the broker to accomplish his aim,
  it should not be considered a failing history, as it does not mean
  that the broker goes out of business.  

  What if the broker receives positive replies from both clients at
  the same time?  It is tempting and perhaps realistic to assume that
  this situation cannot occur, but that assumption amounts to a
  constraint on the environment, and we do not consider in this paper
  such a restriction of the theory to some ``admissible'' histories.
  Furthermore, when the broker really cannot receive two replies
  simultaneously, this will be because of some aspect of his receiving
  mechanism, and it seems that such aspects can be modeled using one of
  the following ideas.

  The broker could systematically prefer one of the clients.  Then a
  history with simultaneous positive replies from both clients would
  be final and successful and would produce updates recording a sale
  to the preferred client.

Alternatively, the broker could break ties non-deterministically.
That is, a history with simultaneous positive replies from both
clients would cause a new query to the environment, asking for a
non-deterministic choice of a client.  Any reply naming a client
would result in a successful final history with a sale to that
client.  

To finish the specification of the algorithm, we must say what is to
be done if netiher client replies, or if one replies negatively and
the other doesn't reply.  One approach is to declare such histories
not to be final.  The broker simply waits, perhaps forever.  

A more realistic approach is that the broker's offers to the clients
include a time limit.  An easy way to model this is to have the empty
history cause, in addition to the offers to the clients, a third query
$t$, to be thought of as a time-out query, i.e., a request to be
informed by the environment, when the time limit has been reached.
Then any history with a reply to this query but no positive reply from
a client will be final and successful and result in no update (or an
update recording that there was no sale and why).  The algorithm can
still hang, if the environment fails to respond to the time-out query,
but that seems a reasonable result for a broker who imposes a
time-limit but can't find out when it has been reached.
\end{ex}

\begin{rmk}
  Although the primary subject of this paper is intra-step interaction
  between an algorithm and its environment, it is not our intention to
  prohibit inter-step interaction of the sort described in \cite{lipari}
  and mentioned in \ref{sec:overview:seqth}.  Inter-step interaction is,
  however, quite easy to describe; the environment is permitted to make
  arbitrary changes to the state's dynamic functions between the steps of
  the algorithm. Thus, a \emph{run} of an algorithm is a (finite or
  infinite) sequence of states, starting with an initial state, in which
  each state but the first is obtained from its predecessor either by an
  intervention of the environment or by a step of the algorithm.  In the
  intervention case, the successor state has the same base set and static
  functions as the predecessor, but the dynamic functions can be
  arbitrarily altered.  In the case of an algorithm step, there is a
  successful final history $\xi$, describing the environment's replies to
  the algorithm's queries during the step, and if the state before the
  step was $X$ then the state after the step is $\tau(X,\xi)$.  Since what
  happens in the intervention steps is quite arbitrary, our concern in
  this paper is to analyze what happens in the algorithmic steps.  The
  notion of run, therefore, occurs only in peripheral remarks like this
  one, not in the main development.

  A referee asked why both intra-step and inter-step interaction are
  needed.  In fact, one could eliminate inter-step interaction, but it
  would require considerable cooperation from the environment because
  of the following complications.  Note that there is no a priori
  bound on the number of locations that the environment could update
  in an inter-step action.  So that action cannot be simulated by a
  small-step algorithm in one step.  This problem can be circumvented
  by having the algorithm ask about these environmental updates only
  when they are needed for the computation.  This means that the
  environment must ``remember'' these updates until they are needed;
  furthermore, it must keep track of whether these updates are
  overwritten by the algorithm's own updates.  \qed\end{rmk}

\subsection{Isomorphism}

As in previous work, starting with \cite{seqth}, we require that the
information relevant to the algorithm's computation that is given by
the state must be explicitly given by the structure of the state, not
implicitly given by the particular nature of its elements.  Formally,
this means that the computation must be invariant under isomorphisms.

Any isomorphism $i:X\cong Y$ between states can be extended in an
obvious, canonical way to act on queries, answer functions, histories,
locations, updates, etc.  We use the same symbol $i$ for all these
extensions.

\begin{unn}{Isomorphism Postulate}
Suppose $X$ is a state and $i:X\cong Y$ is an isomorphism of
$\Upsilon$-structures.  Then:
\begin{ls}
  \item $Y$ is a state, initial if $X$ is.
  \item $i$ preserves causality, that is, if $\xi\vdash_Xq$ then
  $i(\xi)\vdash_Yi(q)$.
  \item $i$ preserves finality, success, and failure, that is, $i(\scr
  F_X^+)=\scr F_Y^+$ and $i(\scr F_X^-)=\scr F_Y^-$.
  \item $i$ preserves updates, that is, $i(\DD(X,\xi))=\DD(Y,i(\xi))$
  for all histories $\xi$ for $X$.
\end{ls}
\end{unn}

\begin{conv}
  In the last part of this postulate, and throughout this paper, we
  adopt the convention that an equation between possibly undefined
  expressions is to be understood as implying that if either side is
  defined then so is the other.
\qed\end{conv}

\begin{rmk}
  We have required that isomorphisms preserve even the irrelevant
  parts of the algorithm, like update sets for unattainable or
  non-final histories.  This requirement could be dropped, thereby
  weakening our postulate, without any damage to our results.
  Nevertheless, it seems a natural requirement in a general
  description of algorithms.  The intuition behind it is that, even if
  an algorithm includes irrelevant information, the states should
  still be abstract; not even the irrelevant information should be
  able to ``see'' the particular nature of the elements of the state.
  For example, if $i:X\to Y$ is an isomorphism, and $\xi$ is an
  unattainable (hence irrelevant) history in $\scr F_X$, then our
  postulate requires $i(\xi)$ (which is equally unattainable and thus
  irrelevant in $Y$) to be in $\scr F_Y$.  To allow the contrary would
  be to allow the algorithm to refer to the difference between $X$ and
  $Y$, a difference that, because of the isomorphism, involves only the
  particular elements constituting these states, not their abstract
  structure. 
  \qed\end{rmk}

\subsection{Small steps}

The final postulate formalizes the requirement that a small-step
algorithm can do only a bounded amount of work in any one step.  The
bound depends only on the algorithm.  Work includes assembling queries
(from elements of the state and labels) and issuing them, computing
what queries to issue, deciding whether the current history suffices
to finish the step, deciding whether the computation has failed, and
computing updates.  Our formalization of this closely follows the
corresponding postulate in \cite[Section~5]{oa1}; the use of a set of
terms to represent the bounded part of the state that the algorithm
looks at goes back to \cite{seqth}.

\begin{unn}{Bounded Work Postulate}
\mbox{}
  \begin{ls}
    \item There is a bound, depending only on the algorithm, for the
    lengths of the tuples in $\Issued_X(\xi)$ , for all states $X$ and
    final, attainable histories $\xi$.
    \item There is a bound, depending only on the algorithm, for the
    cardinality $|\Issued_X(\xi)|$, for all states $X$ and final,
    attainable histories $\xi$.
    \item There is a finite set $W$ of $\Upsilon$-terms (possibly
    involving variables), depending only on the algorithm, with the
    following property.  Suppose $X$ and $X'$ are two states and $\xi$
    is a history for both of them.  Suppose further that each term in
    $W$ has the same value in $X$ as in $X'$ when the variables are
    given the same values in \ran{\ans\xi}.  Then:
    \begin{ls}
      \item If $\xi\vdash_Xq$ then $\xi\vdash_{X'}q$ (so in particular
      $q$ is a query for $X'$).
      \item If $\xi$ is in $\scr F_X^+$ or $\scr F_X^-$, then it is
      also in $\scr F_{X'}^+$ or $\scr F_{X'}^-$, respectively.
      \item $\DD(X,\xi)=\DD(X',\xi)$.
    \end{ls}
  \end{ls}
\end{unn}

This completes our list of postulates, so we are ready to define the
class of algorithms to be treated in this paper and \cite{ga2}.

\begin{df}   \label{alg-def}
  An \emph{interactive, small-step algorithm} is any entity satisfying
  the States, Interaction, Step, Isomorphism, and Bounded Work
  Postulates.
\qed\end{df}

Since these are the only algorithms under consideration in most of
this paper, we often omit ``interactive, small-step''; on the other
hand, when we want to emphasize the difference between these
algorithms and the ``ordinary algorithms'' treated in \cite{oa1, oa2,
  oa3}, we may refer to the present class of algorithms as ``general,
interactive, small-step algorithms''.

As we explained near the beginning of this section,
Definition~\ref{alg-def} provides a formal meaning for ``(interactive,
small-step) algorithm'', whereas the preceding discussion, leading up
to the postulates, was based on an informal notion of (interactive,
small-step) algorithm.  The definition expresses our belief that the
formal notion accurately represents the informal one, i.e., that our
postulates adequately describe the intuitive notion.  

The remainder of this section is devoted to some terminology and
results connected with the Bounded Work Postulate.

\begin{df}
  A set $W$ with the property required in the third part of the
Bounded Work Postulate is called a \emph{bounded exploration witness}
for the algorithm.  Two pairs $(X,\xi)$ and $(X',\xi)$, consisting of
states $X$ and $X'$ and a single $\xi$ that is a history for both, are
said to \emph{agree} on $W$ if, as in the postulate, each term in $W$
has the same value in $X$ as in $X'$ when the variables are given the
same values in \ran{\ans\xi}.
\qed\end{df}

The first two parts of the Bounded Work Postulate assert bounds for
final, attainable histories.  They imply the corresponding bounds for
all attainable histories, thanks to the following lemma.

\begin{la}
Let $X$ be a state.
  \begin{ls}
    \item Any coherent history for $X$ is an initial segment of a
    complete, coherent history for $X$.
    \item Any attainable history for $X$ is an initial segment of a final,
    attainable history for $X$.
  \end{ls}
\end{la}

\begin{proof}
Since $X$ is fixed throughout the proof, we omit it from the
notation.

To prove the first assertion, let $\xi$ be a coherent history.
According to the definition, its order-type (meaning, strictly
speaking, the order-type of the ordering induced by $\leq_\xi$ on the
equivalence classes under $\equiv_\xi$) is some ordinal number
$\alpha$.  We shall inductively define a sequence of coherent
histories $\xi_n$, starting with $\xi_0=\xi$.  Here $n$ will range
over either the set \bbb N of all natural numbers or the set of
natural numbers up to some finite $N$ to be determined during the
construction.  After $\xi_n$ is defined, if it is complete, then stop
the construction, i.e., set $N=n$.  If $\xi_n$ is incomplete, this
means that the set
$D_n=\text{Pending}(\xi_n)=\Issued(\xi_n)-\dom{{\ans\xi}_n}$ is
nonempty.  Extend the answer function $\ans\xi_n$ by adjoining $D_n$
to its domain and assigning it arbitrary values (in $X$) there.  Call
the resulting answer function $\ans\xi_{n+1}$, and make it into a
history $\xi_{n+1}$ by pre-ordering its domain as follows. On
$\dom{{\ans\xi}_n}$, $\leq_{\xi_{n+1}}$ agrees with $\leq_{\xi_n}$.  All
elements of $\dom{\ans\xi_{n+1}}$ are $\leq_{\xi_{n+1}}$ all elements of
$D_n$.  Elements of $D_n$ are $\leq_{\xi_{n+1}}$ only each other.  In
other words, we extend the ordering of $\dom{{\ans\xi}_n}$ by adding $D_n$
at the end, as a single equivalence class.  It is straightforward to
check that each $\xi_n$ (if defined, i.e., if the sequence hasn't
ended before $n$) is a coherent history; the order-type of its
equivalence classes is $\alpha+n$.

If $\xi_n$ is defined for only finitely many $n$, then this is because
the last $\xi_n$ was complete, and so we have the required result.  It
remains to consider the case where $\xi_n$ is defined for all natural
numbers $n$.  In this case, let $\zeta$ be the union of all the
$\xi_n$'s.  (More formally, if we regard functions and orderings as
sets of ordered pairs, then $\ans\zeta$ is the union of the
$\ans\xi_n$'s, and the pre-order $\leq_\zeta$ is the union of the
$\leq_{\xi_n}$'s.)  Then $\zeta$ is also coherent.  Indeed, if
$q\in\dom{\ans\zeta}$ then there is some $n$ such that $q\in\dom{{\ans\xi}_n}$.  As
$\xi_n$ is coherent, $q\in\Issued(\xi_n\restr(<q))$.  But,
since each $\xi_n$ is an initial segment of the next, and therefore of
$\zeta$, we have $\xi_n\restr(<q)=\zeta\restr(<q)$ and so
$q\in\Issued(\zeta\restr(<q))$.  Furthermore, the order-type of
$\zeta$ is $\alpha+\omega$ (where $\omega$ is the order-type of the
natural numbers), so it is well-ordered.

To finish the proof of the first part of the lemma, we need only check
that $\zeta$ is complete.  Suppose, toward a contradiction, that
$q\in\text{Pending}(\zeta)$.  So there is an initial segment $\eta$ of
$\zeta$ such that $\eta\vdash q$.  By the Interaction Postulate,
$\eta$ is finite, so it is an initial segment of $\xi_n$ for some $n$.
Then
$$
q\in\Issued(\xi_n)\subseteq \dom{\ans\xi_{n+1}}\subseteq\dom{\ans\zeta}.
$$
That contradicts the assumption that $q\in\text{Pending}(\zeta)$, so
we have shown that $\zeta$ is complete.  Thus, $\xi=\xi_0$ is an
initial segment of the complete, coherent history $\zeta$, and the
first assertion of the lemma is proved.

To prove the second assertion, let $\xi$ be an attainable history.  In
particular, it is coherent, so, by the first assertion, it is an
initial segment of a complete, coherent history $\zeta$.  By Part~A of
the Step Postulate, $\zeta$ has an initial segment $\eta$ that is a
final history.  If $\zeta$ has several initial segments that are final
histories, then let $\eta$ be the shortest of them; thus no proper
initial segment of $\eta$ is final.  Since both $\xi$ and $\eta$ are
initial segments of $\zeta$, one of them is an initial segment of the
other.  Since $\xi$ is attainable and $\eta$ is final, $\eta$ cannot
be a proper initial segment of $\xi$.  Therefore, $\xi$ is an initial
segment of $\eta$.  Furthermore, $\eta$ is coherent, because it is an
initial segment of the coherent history $\zeta$.  It follows, since no
proper initial segment of $\eta$ is final, that $\eta$ is attainable,
as desired.
\end{proof}

The following corollary extends the first assertion in the Bounded
Work Postulate to histories that need not be final.

\begin{cor}  \label{bd-query}
  There is a bound, depending only on the algorithm, for the lengths
    of the tuples in $\Issued_X(\xi)$ for all states $X$ and all
    attainable histories $\xi$.
\end{cor}

\begin{proof}
  The bound on lengths of queries issued by final, attainable
  histories, given by the Bounded Work Postulate, applies to all
  attainable histories, because these are, by the lemma, initial
  segments of final ones.
\end{proof}

The next corollary similarly extends the second assertion of the
Bounded Work Postulate, and adds some related information.

\begin{cor}  \label{bd-dom}
  There is a bound, depending only on the algorithm, for
    $|\Issued_X(\xi)|$, for all states $X$ and attainable
    histories $\xi$.  The same number also bounds $|\dom{\ans\xi}|$ for all
    attainable $\xi$.
\end{cor}

\begin{proof}
  By the lemma, any attainable history $\xi$ is an initial segment of
  a final, attainable history $\zeta$.  Then $\Issued_X(\xi)
  \subseteq \Issued_X(\zeta)$.  The bound provided by the
  Bounded Work Postulate for $|\Issued_X(\zeta)|$ thus applies
  to $\xi$ as well.  This proves the first assertion of the corollary,
  and the second follows because $\xi$, being attainable, is coherent,
  which implies $\dom{\ans\xi}\subseteq\Issued_X(\xi)$.
\end{proof}

\section{Equivalence of Algorithms}
\label{equiv}

One of our principal aims in this paper and its sequel \cite{ga2} is
to show that every algorithm, in the sense defined above, is
behaviorally equivalent, in a strong sense, to an ASM.  Of course,
this goal presupposes a precise definition of the notion of behavioral
equivalence, and we devote the present section to presenting and
justifying that definition.  As in earlier work on the ASM thesis,
beginning in \cite{seqth} and continuing in \cite{parth, oa1, oa2,
  oa3}, the definition of equivalence is intended to express the idea
that two algorithms behave the same way in all possible situations.
We must, of course, make precise what is meant by ``behave the same
way'' and by ``possible situations.''  Much of what needs to be said
here was already said in \cite[Section~6]{oa1} in the more restricted
context of ordinary interaction, and we shall refer to some of that
discussion in motivating our definition of behavioral equivalence.

Part of the definition of equivalence of algorithms is
straightforward.  As in previous work, we require equivalent
algorithms to have the same states, the same initial states, the same
vocabulary, and the same labels.  The requirement that they agree as
to states and initial states is clearly necessary for any comparison
at all between their behaviors, specifically the aspect of behavior
given by the progression of states in a run.  The requirement that
they agree as to vocabulary is actually a consequence of agreement as
to states (and the requirement, in the definition of algorithm, that
there be at least one state), because any structure determines its
vocabulary.  The requirement that they agree as to labels ensures that
the algorithms have, in any state, the same potential queries; this is
needed for any comparison between their behaviors, since issuing
queries is observable behavior.

The requirements just discussed say that equivalent algorithms agree
as to all the items introduced in the States Postulate.  It is
tempting to go through the remaining postulates, look for statements
of the form ``the algorithm determines'' such-and-such, and require
equivalent algorithms to have the same such-and-such.  Unfortunately,
this approach, which in \cite{seqth} would produce the correct notion
of equivalence, is too restrictive when applied to interactive
algorithms in \cite{oa1} and the present paper.

The difficulties were already pointed out in \cite[Section~6]{oa1} in
connection with the causality relation.  The examples given there
exhibit the following two sorts of problems.  First, a causality
relation could have instances $\xi\vdash_Xq$ whose $\xi$ could never
actually occur in the execution of the algorithm, for example because
$\dom{\ans\xi}$ contains queries that the algorithm would never issue.
Second, as in \cite[Example~6.4]{oa2}, the $\xi$ in an instance
$\xi\vdash_Xq$ of causality could contain redundant elements, such as
a query-reply pair that would have to be present in order for another
query in $\dom{\ans\xi}$ to be issued.  Algorithms whose causality
relations differ only in such irrelevant ways should count as
equivalent.  That is, we should care only about what queries the
algorithm issues in response to histories that can actually occur when
this algorithm runs.

Similar comments apply to the remaining ingredients of an algorithm,
in which histories are used.  The notions of final, successful, and
failing histories and the update sets should not be required to agree
completely when two algorithms are equivalent; it suffices that they
agree on those histories that can actually occur.

\begin{rmk} \label{cpt2} We emphasize again that we are dealing here
  with a situation where only the algorithm and its environment are
  involved.  When several algorithms interact (and we do not choose to
  consider each as a part of the environment for the others), the
  situation would be more complex.  Consider, for example, two
  algorithms that function as components in a larger computation.  The
  first of these components might issue a query whose reply is used by
  the second.  In that case, instances $\xi\vdash_Xq$ of the second
  component's causality relation can be relevant even if that
  component would never issue the queries in $\dom{\ans\xi}$.  Some
  aspects of this situation will arise in \cite{ga2} when we discuss
  the ``do in parallel'' construct of ASMs; a thorough discussion will
  be provided in \cite{composite}.  \qed\end{rmk}

To formalize the preceding discussion, we must still say something
about the notion of a history that can actually appear.  We have
already introduced a precise version of this notion, namely the notion
of an attainable history.  This notion, however, depends on the
causality relation and the notion of finality given with the
algorithm.  When we define equivalence of two algorithms, which one's
attainability should we use?  There are two natural choices: Require
the algorithms to agree (as to queries issued, finality, success,
failure, and updates) on those histories that are attainable for both
algorithms, or require agreement on all histories that are attainable
under at least one of the algorithms.  (There are also some less
natural choices, for example to require agreement on the histories
attainable under the first of the two algorithms; this threatens to
make equivalence unsymmetric.)  Fortunately, these options lead, as we
shall prove below, to the same notion of equivalence of algorithms.
Having advertised our notion of equivalence as a strong one, we take
the apparently stronger of the natural definitions as the official one
and then prove its equivalence with the apparently weaker one.

\begin{df}   \label{equiv-def}
Two algorithms are \emph{behaviorally equivalent} if
\begin{ls}
  \item they have the same states (therefore the same vocabulary), the
  same initial states, and the same labels,
  \item in each state, they have the same attainable histories,
  \item for each state $X$ and each attainable history $\xi$, they
  have the same set $\Issued_X(\xi)$,
  \item for each state, the two algorithms have the same attainable
  histories in $\scr F_X^+$
  and $\scr F_X^-$ (and therefore also in $\scr F_X$).
  \item for each state and each attainable, final, successful
  history, they have the same update sets.
\end{ls}
\qed\end{df}

The following lemma shows that the notion of equivalence is unchanged
if we delete the second item in the definition and weaken the
subsequent ones to apply only to histories that are attainable for
both algorithms.

\begin{la}   \label{equiv-la}
Suppose two algorithms have the following properties.
\begin{ls}
  \item They have the same states (therefore the same vocabulary), the
  same initial states, and the same labels,
  \item for each state $X$ and each history $\xi$ that is attainable
  for both algorithms, they have the same set $\Issued_X(\xi)$,
  \item for each state, any history that is attainable in both
  algorithms and is in $\scr F_X^+$ or $\scr F_X^-$ for one of the
  algorithms is also in $\scr F_X^+$ or $\scr F_X^-$, respectively,
  for the other.
  \item for each state and each history that is attainable for both
  and final and successful (for one and therefore both), they have
  the same update sets.
\end{ls}
Then these two algorithms are equivalent.
\end{la}

\begin{proof}
Comparing the hypotheses of the lemma with the definition of
equivalence, we see that it suffices to prove that, under the
hypotheses of the lemma, any history that is attainable in a state $X$
for either algorithm must also be attainable in $X$ for the other.
Indeed, this would directly establish the second clause of the
definition, and it would make the subsequent clauses in the definition
equivalent to the corresponding hypotheses in the lemma.

Let $A_1$ and $A_2$ be two algorithms satisfying the hypotheses of the
lemma.  Fix a state $X$ for the rest of the proof; we shall suppress
explicit mention of $X$ in our notations.  To prove that $A_1$ and
$A_2$ have the same attainable histories, we first recall that any
attainable history (for either algorithm) has finite domain by
Corollary~\ref{bd-dom} and therefore has finite length.  We can
therefore proceed by induction on the length of histories.  Since the
empty history vacuously satisfies the definitions of ``coherent'' and
``attainable'', we need only verify the induction step.

Consider, therefore, a nonempty history $\xi$ that is attainable for
$A_1$.  Let the last equivalence class in $\dom{\ans\xi}$ be $L$ and let
$\eta$ be the initial segment of $\xi$ obtained by deleting $L$ from
the domain.  Then $\eta$ is attainable for $A_1$ (because $\xi$ is)
and therefore also for $A_2$ (by induction hypothesis).  Furthermore,
the linear order of the equivalence classes of $\xi$, being finite,
is certainly a well-ordering.  So to complete the proof that $\xi$ is
attainable for $A_2$, it suffices to verify that it satisfies, with
respect to $A_2$, the first clause in the definition of coherence
(that every $q\in\dom{\ans\xi}$ is caused by an initial segment that ends
before $q$) and that no proper initial segment of it is final.

For the first of these goals, observe that the induction hypothesis
gives what we need if $q\in\dom{\ans\eta}$, so we need only deal with the
case that $q\in L$.  In this case, we know that
$q\in\Issued^{A_1}(\eta)$ because $\xi$ is coherent for $A_1$.  But
since $\eta$ is attainable for both algorithms, the second hypothesis
of the lemma applies, and we infer that $q\in\Issued^{A_2}(\eta)$, as
required.

For the second goal, we already know that $\eta$ is attainable for
$A_2$ and so none of its proper initial segments can be final.  The
only proper initial segment of $\xi$ not covered by this observation
is $\eta$, so it remains only to show that $\eta$ is not final for
$A_2$.  Since $\eta$ is not final for $A_1$ (because $\xi$ is
attainable for $A_1$) and since $\eta$ is attainable for both
algorithms, the third hypothesis of the lemma immediately gives the
required conclusion.
\end{proof}

A referee asked why equivalence of algorithms is not defined in a more
usual way, by defining a notion of \emph{run} and then defining two
algorithms being equivalent if they have the same runs.  The
alternative definition of equivalence is possible but it is not
simpler.  A run would not be merely a sequence of states as in
\cite{seqth, parth}; it would have to include the queries and replies
as well.  Note also that our definition of equivalence requires that
the two algorithms have the same behavior at every state $X$ whether
$X$ is reachable or not.  It is thus independent of the set of initial
states.  To match this, the run-based definition of equivalence should
allow runs to start at arbitrary states.

\section{Conclusion}

We have formally defined the class of interactive small-step
algorithms (Definition~\ref{alg-def}) by means of postulates that
generalize the postulates of \cite{seqth} and \cite{oa1}, and we have
explained why we believe that this definition matches the intuitive
notion of interactive small-step algorithm.  We have also defined a
strong notion of behavioral equivalence for such algorithms.  This
work provides the essential prerequisites for an analysis of these
algorithms in terms of abstract state machines and for the proof of
thhe ASM thesis in thee small-step interactive case.  That analysis
and proof will be carried out in \cite{ga2}, beginning with the
definition of the appropriate class of ASMs.

\end{document}